\documentclass[reprint,eqsecnum,floats,aps,amsmath,amssymb,nofootinbib,prd,onecolumn, showpacs]{revtex4-1}

\usepackage{graphicx}
\usepackage{amsmath,amssymb}
\usepackage{hyperref}
\usepackage{graphicx}
\usepackage{subfigure}
\usepackage{arydshln}
\usepackage[spanish]{babel}
\usepackage[utf8x]{inputenc}

\begin{document}
	
	\author{Beatriz Elizaga Navascu\'es}
	\email{beatriz.b.elizaga@gravity.fau.de}
	\affiliation{Institute for Quantum Gravity, Friedrich-Alexander University Erlangen-N{\"u}rnberg, Staudstra{\ss}e 7, 91058 Erlangen, Germany}
	\author{Guillermo A. Mena Marug\'an}
	\email{mena@iem.cfmac.csic.es}
	\affiliation{Instituto de Estructura de la Materia, IEM-CSIC, Serrano 121, 28006 Madrid, Spain}
	\author{Santiago Prado}
	\email{santiago.prado@iem.cfmac.csic.es}
	\affiliation{Instituto de Estructura de la Materia, IEM-CSIC, Serrano 121, 28006 Madrid, Spain}

\title{Asymptotic diagonalization of the fermionic Hamiltonian in hybrid loop quantum cosmology} 

\begin{abstract} 

We use the freedom available in hybrid loop quantum cosmology to split the degrees of freedom between the geometry and the matter fields so as to build a quantum field theory for the matter content with good quantum properties. We investigate this issue in an inflationary, flat cosmology with inhomogeneous perturbations, and focus the discussion on a Dirac field, minimally coupled to the cosmological background and treated as a perturbation. After truncating the action at the lowest nontrivial order in perturbations, one must define canonical variables for the matter content, for which one generally employs canonical transformations that mix the homogeneous background and the perturbations. Each of these possible definitions comes associated with a different matter contribution to the Hamiltonian of the complete system, that may, in general, contain terms that are quadratic in creationlike variables, and in annihilationlike variables, with the subsequent production and destruction of pairs of fermionic particles and antiparticles. We determine a choice of the fermionic canonical variables for which the interaction part of the Hamiltonian can be made as negligible as desired in the asymptotic regime of large particle/antiparticle wave numbers. Finally, we study the quantum dynamics for this choice, imposing the total Hamiltonian constraint on the quantum states and assuming that their gravitational part is not affected significantly by the presence of fermions. In this way, we obtain a Schrödinger equation for the fermionic degrees of freedom in terms of quantum expectation values of the geometry that leads to asymptotically diagonal Heisenberg relations and Bogoliubov evolution transformations, with no divergences in the associated normal-ordered Hamiltonian.

\end{abstract}


\maketitle 

\newpage

\section{Introduction}

Choosing a Fock representation for the quantization of matter fields in curved spacetimes, and with it a vacuum state, is a nontrivial task even in the case of linear fields. In quantum mechanics, when one is considering systems with finite degrees of freedom, one can make use of results like the Stone-von Neumann theorem that guarantees that there exists only one representation of the Weyl relations, up to unitary equivalence, with the desired properties, namely a strongly continuous, irreducible, and unitary representation \cite{Stone, VN}. Nevertheless, when one has to deal with fields, that are systems with infinite degrees of freedom, there is no such theorem at our disposal and, in the best of cases, one has to appeal to symmetries or other kinds of physical arguments in order to select a vacuum. In Minkowski spacetime, for instance, the most natural thing to ask for is Poincar\'e invariance, which in fact picks out a unique representation, up to unitary transformations \cite{Peskin,haag}. For stationary spacetimes, the so-called energy criterion can be used to select a preferred complex structure (which essentially fixes the Fock representation) out of the infinite many that are possible \cite{energy1,energy2}. In spite of all this, no general uniqueness result has been found for systems with fieldlike degrees of freedom in nonstationary spacetimes, such as cosmologies \cite{bidav,Wald}.  Actually, in nonstationary spacetimes, and after imposing invariance under the spatial isometries that the system possesses, one could expect that the ambiguity that affects the choice of representation could be solved, not by demanding invariance under time evolution, since the dynamics is not a symmetry anymore, but by requiring that, at least, the quantum evolution of the creation and annihilation operators can be implemented in a unitary way. For a variety of cosmological systems, it has been recently proven that this criterion of unitarity (together with the invariance under spatial symmetries) indeed determines a preferred family of vacua, which are all unitarily related \cite{uniqueness1,uniqueness2,uniqueness3,uniqueness4,uniqueness5,uniqueness6,uniqueness7,uniqueness8,uniqueness9,uniqueness10,uniqueness11}.

The system that we study in this work is a flat Friedmann-Lema\^itre-Robertson-Walker (FLRW) cosmology with a Dirac field that will be regarded as a perturbation, including its zero mode if it is not identically vanishing. Since we are not concerned here about infrared divergences, in order to simplify the discussion and keep all definitions rigorous we restrict our considerations to compact FLRW spatial sections. More specifically, we analyze the case of sections with the topology of a three-torus. In this system, it has been proven that there is a unique family of Fock representations of the Dirac field, all related among them by unitary transformations, for which the vacuum is invariant under the isometries of the spatial sections and the Heisenberg dynamics of the creation and annihilation operators is unitarily implementable (once one adopts a standard convention for particles and antiparticles), provided that one treats the FLRW spacetime as a classical or {\sl effective} background \cite{uniqfermi}.

Furthermore, the system was further studied within the hybrid approach for the quantization of gravitational models, in which the (matter) fields are quantized with suitable Fock representations and the homogeneous geometry is quantized with techniques inspired by nonperturbative quantum gravity, typically with methods of the canonical formalism known as loop quantum gravity (LQG) \cite{Thiemann}. This hybrid approach has been successfully used in cosmological scenarios with perturbations \cite{hlqc1,hlqc2,hlqc3}. In our particular system, we treat the degrees of freedom of the homogeneous cosmology exactly and truncate the action at quadratic order in the perturbations, that is the lowest order with a nonvanishing contribution. With this truncation, the zero mode of the Hamiltonian constraint is formally equal to the constraint of the homogeneous cosmology plus a contribution that is quadratic in the perturbations. Nevertheless, it is worth pointing out that there is an inherent freedom in the way in which one decides to separate the degrees of freedom of the homogeneous cosmology from the inhomogeneous perturbations, since they can always be remixed using transformations that preserve the canonical symplectic structure at the level of the perturbative truncation of the system. Actually, instead of considering this freedom a nuisance, the idea that was put forward in Ref. \cite{backreaction} was to exploit the freedom to define canonical variables for which the Hamiltonian of the perturbative Dirac variables had certain nice properties. In fact, when fermions were first studied within the hybrid approach in Ref. \cite{fermihlqc}, the choice of fermionic variables was based only on the requirements of invariance of the resulting Fock vacuum under the spatial symmetries, a unitarily implementable Heisenberg evolution in the regime of quantum field theory in curved spacetimes, and a standard convention for particles and antiparticles. But it was already shown there that ultraviolet divergences appeared in the resulting Schrödinger equation for the fermionic degrees of freedom (after a convenient sort of Born-Oppenheimer approximation in the imposition of the full quantum constraint). These divergences could only be solved by either a regularization scheme with {\sl substraction of infinities} or by introducing a further restriction on the choice of perturbative variables that define the vacuum \cite{backreaction}. In practice, this new restriction lowered the asymptotic order of the interaction part of the fermionic Hamiltonian at large wave numbers (identified as the eigenvalues of the Dirac operator on the spatial sections), diminishing the production of pairs of particles and antiparticles in this asymptotic regime.

In the present work, go one step beyond in the same direction and, by further taking advantage of the freedom to split the degrees of freedom between the geometry and the perturbations, prove that one can absorb the interaction terms of the fermionic Hamiltonian so as to make them as negligible as desired in the asymptotic regime of large eigenvalues (in absolute value) of the modes of the Dirac field. Moreover, in this way we not only improve the quantum behavior of the fermionic contribution to the Hamiltonian of our gravitational model, but we also reduce the ambiguity in our choice of Fock representation and the vacuum for the Dirac field, leaving only some remaining asymptotic freedom in certain phases. In addition, we also notice that, since the resulting fermionic Hamiltonian contribution is diagonal, at least asymptotically, the dynamics that it generates is very simple for the vacuum of the representation, essentially a rotating phase. In this sense, one can think of this vacuum and our splitting of degrees of freedom in the hybrid quantization as those that are best adapted to the dynamics of the entire cosmological system.

The rest of the paper is organized as follows, in Sec. \ref{model}, and following the procedure detailed in Ref. \cite{fermihlqc}, we introduce the Dirac field as a perturbation around the flat FLRW cosmology, truncating the action at quadratic perturbative order. In Sec. \ref{fermiH} we consider a generic choice of creation and annihilationlike variables for the Dirac field, allowing definitions that depend on the homogeneous geometry. This dependence is captured in the coefficients of the linear transformations that relate our fermionic variables with the coefficients of the fermionic mode expansions. We also calculate the form of the fermionic contribution to the total Hamiltonian constraint for each of the possible choices of variables, paying special attention to the nondiagonal part of this contribution, that provides the fermionic interaction terms. Then, in Sec. \ref{diag} we adopt an {\sl ansatz} for the creation and annihilationlike variables inspired in the analysis of Refs. \cite{fermihlqc, backreaction}, and we investigate the specific expression that these variables must take so that they asymptotically diagonalize the fermionic Hamiltonian. We also give the form of the remaining, diagonal part of the Hamiltonian. Finally, in Sec. \ref{HQ} we briefly revisit the hybrid quantization of Ref. \cite{fermihlqc} to adapt it to our new variables. In particular, we give the new Schr\"odinger equation, Heisenberg relations, Bogoliubov evolution transformation, and evolution operator for the new choice of vacuum. We conclude in Sec. \ref{conclusiones}, summarizing our results, and commenting on some lines for further research.

\section{Classical Model}\label{model}
   
Let us start by briefly presenting the model that we study. The homogeneous sector consists of a flat FLRW spacetime with compact spatial sections that are isomorphic to the three-torus, $T^3$, and of a massive scalar field, $\phi$, subject to a potential $V(\phi)$. This field plays the role of the inflaton. We use spacetime coordinates that exploit the symmetry of this cosmological background. The inhomogeneous sector is given by a Dirac field with mass $M$, which is treated as a perturbation. For all practical purposes, we include in this sector also the zero mode of the Dirac field if it does not vanish, regarding it as a perturbative degree of freedom. We then truncate the action at quadratic order in the perturbations \cite{Death, Hawking}, and use the canonical structure and the Hamiltonian of this truncated action to construct our description of the entire system. 

In principle, we can also add perturbative inhomogeneities to the metric and the inflaton, as in Refs. \cite{CMM,tensor}, perturbations that we consider again as part of the inhomogeneous sector. These additional perturbations originate new quadratic contributions to the purely homogeneous part of the Hamiltonian constraint, and they furthermore introduce a whole family of linear perturbative constraints. The only perturbative quantities that are physically meaningful are those that commute with this family of constraints, and they are generally called gauge-invariant perturbations \cite{Bardeen,Mukhanov}. Invariant perturbations of this kind, for the case with flat spatial topology that we are discussing, are the tensor perturbations of the metric and the Mukhanov-Sasaki scalar \cite{MS,Sasaki,kodasasa}, which mixes scalar perturbations of the metric and the inflaton. A phase space for the perturbations can then be constructed with these gauge invariants and with an Abelianized version of the perturbative linear constraints, together with suitable canonical momenta of all of them. Nonetheless, since the definitions of these variables make use of the homogeneous FLRW ones, they do not form a canonical set with the variables of the homogeneous sector, and these latter variables have to be modified with quadratic terms in the perturbations in order to render the whole set canonical again. On the other hand, since the Einstein-Dirac action is quadratic in the Dirac field, the fermionic perturbations, at our order of truncation, couple directly to the homogeneous tetrad and hence turn out to be gauge invariants (namely, they commute under Poisson brackets with the linear perturbative constraints arising from the perturbation of the tetrad).

It is convenient to rescale the Dirac field by a factor $e^{3\tilde \alpha/2}$ in order to get canonical Dirac brackets that are constant in the evolution, after imposing an internal time gauge on the homogeneous tetrad \cite{TN}. Here, $\tilde \alpha$ is the logarithm of the scale factor of the FLRW geometry up to an additive constant $\ln[4\pi/(3l_0)^3]/2$, where $l_0$ is the compactification length of the tori, and, in general, the tilde over a homogeneous variable indicates that it has been corrected with quadratic perturbative terms, as we have mentioned above. In the adopted internal time gauge, we can expand the two chiral components of the rescaled field in modes of the Dirac operator on the spatial sections. This expansion is especially suitable because the Dirac operator is invariant under the spatial isometries of the FLRW cosmology, a property that at the end of the day guarantees that the dynamical equations do not mix its eigenmodes. Since we are dealing with compact spatial sections, the spectrum of this operator is discrete. The eigenvalues for $T^3$ are $\pm \omega_k=\pm 2\pi\vert\vec{k}+\vec{\tau}\vert/l_0$, $\vec{k}\in\mathbb{Z}^3$, and where $\vec{\tau}=\sum\theta_I \vec{v}_I/2$ characterizes the spin structure on the spatial sections ($\theta_I=0$ or $1$ depending on the spin structure, and $\vec{v}_I$ is the standard $\mathbb{Z}^3$ basis). Then, the rescaled Dirac field can be described in terms of a set of time-dependent Grassmann variables $\{m_{\vec{k}}.\bar r_{\vec{k}},\bar s_{\vec{k}}, t_{\vec{k}}\}$, where the bar denotes complex conjugation. The ordered pairs $(m_{\vec{k}},\bar r_{\vec{k}})$ and $ (s_{\vec{k}},\bar t_{\vec{k}})$ are simply the coefficients of the Dirac eigenspinors of the left-handed and (the complex conjugate of the) right-handed components of the rescaled Dirac field, respectively, up to a multiplicative constant $[4\pi/(3l_0)]^{-3/4}$. The first variable of each of these pairs is associated with the eigenspinors that have positive eigenvalues, while the second variable corresponds to negative eigenvalues. Each of these mode coefficients is canonically conjugate to its complex conjugate, inasmuch as their Dirac bracket equals $-i$, whereas the rest of the Dirac brackets between our fermionic variables vanish \cite{TN}.
  
\section{Fermionic Contribution to the Hamiltonian}\label{fermiH}

Let us introduce the following family of annihilationlike variables of particles and creationlike variables of antiparticles for the Dirac field, respectively defined by these linear combinations of the Grassman variables that determine the field,
\begin{align}\label{varicre}
  a^{(x,y)}_{\vec{k}}=f_1^{\vec{k},(x,y)} x_{\vec{k}}+f_2^{\vec{k},(x,y)} \bar{y}_{-\vec{k}-2\vec{\tau}}, \quad  \bar{b}^{(x,y)}_{\vec{k}}=g_1^{\vec{k},(x,y)} x_{\vec{k}} +g_2^{\vec{k},(x,y)} \bar{y}_{-\vec{k}-2\vec{\tau}}.
\end{align}
Here, $(x_{\vec{k}},\bar y_{\vec{k}})$ is any of the ordered pairs $(m_{\vec{k}}.\bar s_{\vec{k}})$ and $ (t_{\vec{k}},\bar r_{\vec{k}})$, and the superindex $(x,y)$ means that the coefficients may be different for each of the pairs. Notice that, in the linear combinations that provide our creation and annihilationlike variables, we have imposed that they do not mix contributions from different modes of the spatial Dirac operator, labeled by the value of $\vec{k}$, so that our definitions respect the spatial symmetries of the fermionic dynamics (and hence
the resulting complex structure is invariant under those symmetries) \cite{uniqfermi}. Since the variables given in Eq. \eqref{varicre} have to satisfy standard anticommutation relations, the coefficients of the linear combinations that define them must fulfil the relations \cite{uniqueness2}
\begin{align} \label{relacio}
f_2^{\vec{k},(x,y)}=e^{i F_2^{\vec{k},(x,y)} }\sqrt{1-\left\vert f_1^{\vec{k},(x,y)} \right\vert^2},\quad g_1^{\vec{k},(x,y)}= e^{iJ^{(x,y)}_{\vec{k}}}\bar{f}^{\vec{k},(x,y)}_2,\quad g_2^{\vec{k},(x,y)}=- e^{iJ^{(x,y)}_{\vec{k}}} \bar{f}^{\vec{k},(x,y)}_1,\end{align}
where $F_2^{\vec{k},(x,y)}$ and $J^{(x,y)}_{\vec{k}}$ are real phases.

In general, we allow linear combinations that depend on the variables of the homogeneous sector, namely  $f_L^{\vec{k},(x,y)}\equiv f_L^{\vec{k},(x,y)}(\tilde\alpha,\pi_{\tilde\alpha},\tilde\phi,\pi_{\tilde\phi})$ and $g_L^{\vec{k},(x,y)}\equiv g_L^{\vec{k},(x,y)}(\tilde\alpha,\pi_{\tilde\alpha},\tilde\phi,\pi_{\tilde\phi})$, where $L=1,2$ and $\pi_{\tilde{z}}$ is the canonical momentum of the variable $\tilde{z}=\tilde\alpha,\tilde\phi$. As a result, the new fermionic variables do not constitute a canonical set with $(\tilde\alpha,\pi_{\tilde\alpha},\tilde\phi,\pi_{\tilde\phi})$, a fact that calls for a suitable redefinition of the homogeneous variables, which must be corrected with quadratic perturbative terms along the lines that we have already explained in order to arrive at a new canonical set ($\alpha,\pi_\alpha,\phi,\pi_\phi)$. At our order of truncation, the desired corrections are \cite{fermihlqc} 
\begin{eqnarray}
\label{newscale}
z-\tilde{z}\equiv \Delta z&=&\frac{i}{2}\sum_{\vec{k},(x,y)}[
(\partial_{\pi_{\tilde{z}}}x_{\vec{k}}) {\bar x}_{\vec{k}}+(\partial_{\pi_{\tilde{z}}}{\bar x}_{\vec{k}}) x_{\vec{k}}+(\partial_{\pi_{\tilde{z}}}y_{\vec{k}}) {\bar y}_{\vec{k}}+(\partial_{\pi_{\tilde{z}}}{\bar y}_{\vec{k}}) y_{\vec{k}} ],
\\
\label{newpi}
{\pi}_{ z}-\pi_{\tilde{z}}\equiv \Delta \pi_ z&=&-\frac{i}{2}\sum_{\vec{k},(x,y)}[
(\partial_{\tilde{z}}x_{\vec{k}}) {\bar x}_{\vec{k}}+(\partial_{\tilde{z}}{\bar x}_{\vec{k}}) x_{\vec{k}}+(\partial_{\tilde{z}}y_{\vec{k}}) {\bar y}_{\vec{k}}+(\partial_{\tilde{z}}{\bar y}_{\vec{k}}) y_{\vec{k}} ],
\end{eqnarray} 
where $z=\alpha,\phi$ and the subindex $(x,y)$ indicates that we are summing over both existing pairs. This change of variables gives rise then to alterations in the homogeneous part of the Hamiltonian constraint, producing new perturbative contributions from it. Truncating those contributions at the relevant perturbative order, one can see that the final result is a new fermionic contribution $\breve{H}_D$ to the zero mode of the Hamiltonian constraint, given by the expression \cite{fermihlqc}
\begin{equation}
\breve{H}_D=H_D-\partial_\alpha H_{\vert 0}\Delta\alpha-\partial_{\pi_\alpha}H_{\vert 0}\Delta\pi_\alpha  -\partial_\phi H_{\vert 0}\Delta\phi-\partial_{\pi_\phi} H_{\vert 0}\Delta\pi_\phi
\end{equation} where $H_D$ is the old fermionic contribution and $H_{\vert 0}$ is the Hamiltonian of the unperturbed model, with their dependence on the old homogeneous variables identified with the new ones.

Following calculations similar to those of Ref  \cite{backreaction}\footnote{As in that reference, we now ignore the possible zero mode of the Dirac field, assuming that we can find a suitable representation for it. This is not important in our discussion, because we focus it on the ultraviolet sector of the field.}, we then obtain the fermionic contribution
\begin{align}
\breve{H}_D&=\sum_{\vec{k},(x,y)}\Bigg[h_D^{\vec{k},(x,y)}\left(\bar{a}^{(x,y)}_{\vec{k}}a^{(x,y)}_{\vec{k}}-a^{(x,y)}_{\vec{k}}\bar{a}^{(x,y)}_{\vec{k}}+\bar{b}^{(x,y)}_{\vec{k}}b^{(x,y)}_{\vec{k}}-b^{(x,y)}_{\vec{k}}\bar{b}^{(x,y)}_{\vec{k}}\right)\nonumber \\
&+h_J^{\vec{k},(x,y)}\left(\bar{b}^{(x,y)}_{\vec{k}}b^{(x,y)}_{\vec{k}}-b^{(x,y)}_{\vec{k}}\bar{b}^{(x,y)}_{\vec{k}}\right) +\bar{h}_I^{\vec{k},(x,y)}\left(a^{(x,y)}_{\vec{k}}b^{(x,y)}_{\vec{k}}\right)-h_I^{\vec{k},(x,y)}\left(\bar{a}^{(x,y)}_{\vec{k}}\bar{b}^{(x,y)}_{\vec{k}}\right)\Bigg],
\end{align} where 
\begin{align}
&h_D^{\vec{k},(x,y)}=\frac{\omega_k}{2e^{\alpha}}\left(\left\vert f_2^{\vec{k},(x,y)}\right\vert ^2-\left\vert f_1^{\vec{k},(x,y)} \right\vert ^2\right) + \tilde  M\Re\left(f_1^{\vec{k},(x,y)}\bar{f}^{\vec{k},(x,y)}_2\right)-\frac{i}{2}\left(\bar{f}^{\vec{k},(x,y)}_1 \partial f_1^{\vec{k},(x,y)}+\bar{f}^{\vec{k},(x,y)}_2 \partial f_2^{\vec{k},(x,y)}\right), \\
&h_J^{\vec{k},(x,y)}=-\frac{1}{2}\partial J^{(x,y)}_{\vec{k}}, \\
&h_I^{\vec{k},(x,y)}=e^{-iJ^{(x,y)}_{\vec{k}}}\bigg[if_2^{\vec{k},(x,y)}\partial f_1^{\vec{k},(x,y)}-if_1^{\vec{k},(x,y)} \partial f_2^{\vec{k},(x,y)}+\frac{2\omega_k}{e^{\alpha}}f_1^{\vec{k},(x,y)}f_2^{\vec{k},(x,y)}+\tilde M\left(f_1^{\vec{k},(x,y)}\right)^2- \tilde M\left(f_2^{\vec{k},(x,y)}\right)^2\bigg].
\end{align}
Here, we have introduced the rescaled mass $\tilde M=2 M \sqrt{\pi/(3l_0^3)}$, we have used $\Re$ to denote the real part of a complex number, and we have defined $\partial$ as the linear differential operator $\partial\equiv\{  H_{\vert 0}  , \ \boldsymbol{\cdot}\ \}$, where $\{ \boldsymbol{\cdot},\boldsymbol{\cdot}\}$ are the Poisson brackets of our truncated system.

\section{Diagonalization of the Fermionic Contribution}\label{diag}

We know from Refs. \cite{backreaction,fermihlqc} that, on the sector of large $\omega_k$, one can lower the asymptotic order of $h_I^{\vec{k},(x,y)}$ with a suitable definition of creation and annihilationlike variables, restricting also in this way the freedom available in their choice. If one were to continue with this procedure, lowering more and more the asymptotic order, one would restrict more and more the choice of fermionic variables, with the hope that one would arrive at a unique choice, perhaps up to certain phases, in the limit of a complete asymptotic diagonalization of $\breve H_D$. Inspired by the analysis of Refs. \cite{backreaction,fermihlqc}, we adopt for $f_1^{\vec{k},(x,y)}$ the following asymptotic series expansion in inverse powers of $\omega_k$: 
\begin{equation}\label{eq: ansatz1}
f_1^{\vec{k},(x,y)}=e^{iF_2^{\vec{k},(x,y)}}\sum^\infty_{n=1}\frac{(-i)^{n+1}\gamma_n}{\omega_k ^n},\end{equation}
with $\gamma_n\in \mathbb{R}$. From the first relation in Eq. \eqref{relacio}, it then follows that, asymptotically,
\begin{equation} \label{eq: ansatz2}
f_2^{\vec{k},(x,y)}=e^{iF_2^{\vec{k},(x,y)}}\sum^\infty_{n=0}\frac{(-i)^n\tilde\gamma_n}{\omega_k ^n},
\end{equation}
with $\tilde\gamma_n\in \mathbb{R}$ again, and
where the coefficients $\tilde\gamma_n$ are defined as
\begin{equation}\label{tildegamma}
\tilde\gamma_0=1, \quad \tilde\gamma_{2n-1}=0,\quad \tilde\gamma_{2n}=(-1)^{n+1}\left[\frac{1}{2}\Gamma_{2n}+\sum^\infty_{m=2}\frac{(2m-3)!!}{(2^m) m!}\sum^{n}_{i_{m-1}=1}...\sum^{i_2}_{{i_1}=1}\Gamma_{2n-2i_{m-1}}...\Gamma_{2i_1}\right] ,\quad \forall n \geq 1 \end{equation}
with
\begin{equation}
\Gamma_0=0,\quad\Gamma_{2n}=\sum^{2n}_{i=1}(-1)^{n+i}\gamma_i \gamma_{2n-i}, \quad  \forall n\geq 1,
\end{equation}
and we set $\gamma_0=0$.

Substituting these series in the formula for the interaction coefficient $h_I^{\vec{k},(x,y)}$ of $\breve H_D$, one finds that
\begin{align}
 h_I^{\vec{k},(x,y)}=e^{-i\left(J^{(x,y)}_{\vec{k}}-2F_2^{\vec{k},(x,y)}\right)}\sum_{n=0}^\infty \left(\frac{-i}{\omega_k }\right)^n A_{n,0},\end{align}
 where
\begin{align}\label{Anm}
A_{n,m}&=\sum_{l=m}^n\Big[\tilde\gamma_{n-l}\partial\gamma_l-\gamma_l\partial \tilde\gamma_{n-l}
-2e^{-\alpha}\tilde\gamma_l\gamma_{n+1-l}-\tilde M(\gamma_l \gamma_{n-l}+\tilde\gamma_l\tilde\gamma_{n-l})\Big],\quad n\geq m, \\
 A_{n,m}&=0, \quad n<m.
\end{align}
It follows that an asymptotic diagonalization of the fermionic contribution to the zero mode of the Hamiltonian constraint is achieved if and only  if $A_{n,0}=0$ for all $n\geq 0$. On the other hand, notice that, if we keep $n\geq 0$ in the formula that gives the coefficients $A_{n,m}$, namely Eq. \eqref{Anm}, $\gamma_{n+1}$ only appears in the case with $m=0$, when one evaluates the contribution corresponding to vanishing label $l$. In addition, the sum that determines $A_{n,0}$ in Eq. \eqref{Anm} can be rewritten as the mentioned contribution with label $l=0$ plus $A_{n,1}$. With these indications, one can easily deduce that $A_{n,0}=0$ implies  the recursive relation
\begin{align}\label{eq: recursive}
\gamma_{n+1}=-\frac{\tilde Me^{\alpha}}{2}\tilde\gamma_n+ \frac{e^\alpha}{2}A_{n,1}.
\end{align}

Since $\tilde\gamma_0=1$, the above relation gives us, in particular, the first term of the series \eqref{eq: ansatz1}, $\gamma_1=-\frac{1}{2}\tilde Me^{\alpha}$, from which we can univocally obtain the rest of the unknown terms. This is possible because $\tilde{\gamma}_n$ is completely determined via Eq. \eqref{tildegamma} by $\gamma_m$ with $m\leq n$, and the nonvanishing coefficients $A_{n,1}$ only involve contributions of $\tilde\gamma_{m}$ with $m\leq n$. For instance, a straightforward calculation shows that $\gamma_2=-\frac{1}{4}e^{-\alpha }\tilde M \pi_\alpha$. Thus \begin{equation}
f_1^{\vec{k},(x,y)}=e^{iF_2^{\vec{k},(x,y)}}\frac{\tilde Me^{\alpha}}{2\omega_k}-ie^{iF_2^{\vec{k},(x,y)}}\frac{\tilde M e^{-\alpha}}{4\omega_k^2} \pi_\alpha+\mathcal{O}\left(\omega_k^{-3}\right),
\end{equation} an equation that of course is consistent with the previous results of Refs. \cite{uniqfermi} and \cite{backreaction}.

The asymptotic form of $f_1^{\vec{k},(x,y)}$ severely restricts our choice of creation and annihilationlike variables, leaving all the asymptotic freedom just in the phases $F_2^{\vec{k},(x,y)}$ and $J^{(x,y)}_{\vec{k}}$. With this choice, or rather with this iterative family of choices, we can make the interaction terms $h^{\vec{k},(x,y)}_I$ vanish in the fermionic Hamiltonian at any desired asymptotic order in inverse powers of $\omega_k$. 

Let us consider now the rest of fermionic contributions to the zero mode of the Hamiltonian constraint, namely the diagonal fermionic terms. It is convenient to define 
\begin{equation}
\tilde f_1^{\vec{k},(x,y)}= e^{-iF_2^{\vec{k},(x,y)}}f_1^{\vec{k},(x,y)}, \quad \tilde f_2^{\vec{k},(x,y)}=e^{-iF_2^{\vec{k},(x,y)}}f_2^{\vec{k},(x,y)},
  \end{equation}
so that
\begin{equation}
h_D^{\vec{k},(x,y)}=\Re\Bigg[\frac{e^{-\alpha}\omega_k}{2}\left(\left\vert \tilde f_2^{\vec{k},(x,y)}\right\vert ^2-\left\vert \tilde f_1^{\vec{k},(x,y)} \right\vert ^2\right) + \tilde M\tilde f_1^{\vec{k},(x,y)}\bar{\tilde f}_2-\frac{i}{2}\left(\bar{\tilde f}_1 \partial \tilde f_1^{\vec{k},(x,y)}+\bar{\tilde f}_2 \partial \tilde f_2^{\vec{k},(x,y)}\right)\Bigg]+\frac{1}{2}\partial F_2^{\vec{k},(x,y)}.
\end{equation}
Here, we have used that the functions $h_D^{\vec{k},(x,y)}$ are always real.\footnote{The only term for which this is not obvious is $-i\left(f_1^{\vec{k},(x,y)}\partial\bar f_1^{\vec{k},(x,y)}+f_2^{\vec{k},(x,y)} \partial \bar f_2^{\vec{k},(x,y)}\right)$. But since $\left\vert f_1^{\vec{k},(x,y)} \right\vert ^2+\left\vert f_2^{\vec{k},(x,y)} \right\vert^2=1$, the term in parenthesis is indeed imaginary.} In what follows, we restrict the complex phase $F_2^{\vec{k},(x,y)}$ so that these functions (which provide diagonal fermionic contributions to the Hamiltonian constraint) do not depend on the momentum $\pi_\phi$ of the homogeneous inflaton. Notice that the only part in this constraint that contains the coefficients $\gamma_n$ and $ \tilde\gamma_n$ is given by
\begin{equation}
\breve h_D^{\vec{k},(x,y)}=h_D^{\vec{k},(x,y)}-\frac{1}{2}\partial F_2^{\vec{k},(x,y)},
\end{equation}
which, with our asymptotic expansions, takes the specific form
\begin{equation}
\breve h_D^{\vec{k},(x,y)} =\sum_{n=-1}^\infty \frac{\breve \gamma_n}{\omega_k^n}.\end{equation}
Finally, a direct calculation shows that the coefficients $\breve \gamma_n$ turn out to be 
\begin{align}    
\breve\gamma_{-1} &= \frac{e^{-\alpha}}{2}, \\
\breve\gamma_n =&\Re(i^{n+1})\sum^n_{l=0}\left[\frac{(-1)^l}{2}\left(2\tilde M\tilde\gamma_l \gamma_{n-l}-2e^{-\alpha}\gamma_{n+1-l}\gamma_l+\gamma_l\partial \gamma_{n-l}+\tilde\gamma_l\partial \tilde\gamma_{n-l}\right)\right],\ \forall n\geq 0.
\end{align}

\section{Hybrid Quantization}\label{HQ}

The results of the previous sections restrict in a physically appealing way the choice of canonical variables for the homogeneous and fermionic parts of the phase space of our truncated cosmology. These canonical variables are the homogeneous pairs $(\alpha,\pi_\alpha)$ and $(\phi,\pi_\phi)$, and the fermionic annihilation and creationlike variables for particles, $\{(a_{\vec{k}}^{(x,y)},\bar{a}_{\vec{k}}^{(x,y)})\}_{\vec{k}\neq0}$, and for antiparticles $\{(b_{\vec{k}}^{(x,y)},\bar{b}_{\vec{k}}^{(x,y)})\}_{\vec{k}\neq0}$, all of them determined by relations \eqref{varicre}-\eqref{newpi}. In particular, the homogeneous variables have been defined so that they commute under Poisson brackets with the fermionic variables at our order of perturbative truncation. We have seen that, with asymptotic expansions of the form \eqref{eq: ansatz1} and \eqref{eq: ansatz2}, the interaction terms in the fermionic contribution to the zero mode of the Hamiltonian constraint can be rendered as negligible as desired in the ultraviolet sector of large wave numbers. Then, if one decides to restrict all considerations to the context of a fermionic field in linearized cosmology, namely if one ignores the quadratic fermionic backreaction on the classical dynamics of the homogeneous variables, the evolution of the introduced annihilation and creationlike variables becomes asymptotically diagonal. In this respect, the important result is that this diagonalization serves as a valid criterion to select canonical variables for the fermionic degrees of freedom, characterizing them on the entire phase space of our cosmological system. Remarkably, nonetheless, the benefits of using this type of fermionic annihilation and creationlike variables lie beyond the linearized context that we have just commented on. Indeed, their definition is compatible with that of the homogeneous variables in our search for a canonical set, and the resulting expression for the Hamiltonian constraint (at the considered truncation order) asymptotically displays only quadratic combinations of these fermionic variables in a way that is proportional to the number operator, once one adopts a Fock representation with normal ordering. This fact simplifies enormously the task of finding a quantum representation of the constraint operator not just for the fermions, but for the combined system that includes the homogenous variables within the framework of hybrid quantum cosmology. Moreover, as it was argued in Refs. \cite{fermihlqc,backreaction} and we discuss in this section, there exist quantum states of the entire cosmology such that the resolution of the quantum constraint in a sort of Born-Oppenheimer approximation amounts to the condition that the fermionic part of their wave functions solves a Schr\"odinger equation. This is similar to the situation found in linearized cosmology, with the very important difference that in our treatment the homogeneous background does not need to correspond to a classical solution or even to an effective trajectory. Rather, the dependence of the linear fermionic equations on this background is given by expectation values of geometric operators in the part of the wave function that describes the homogeneous geometry and, as we have pointed out, in principle it is not  necessary that these expectation values evolve as in general relativity or according to any effective dynamics.

In order to proceed to the hybrid quantization of the system, with its phase space already split into different sectors in the way that we have explained, we choose suitable representations for each of those sectors and represent the total system on the tensor product of the partial representation spaces. For the FLRW geometry, corrected with quadratic perturbative contributions according to our comments, we pick out a representation inspired in LQG and specified in Ref. \cite{mmo} (see also Refs. \cite{aps1,aps2}). Therefore, instead of using the canonical variables $\{\alpha,\pi_\alpha\}$ we employ the alternative variable $v$, proportional to the physical volume of the spatial sections of our cosmological model, together with its canonical momentum $b/2$, which is proportional to the Hubble parameter \cite{AshSingh}. More specifically, we have that $|v|= [16\pi/(27 l_0^3 \gamma^2 \Delta_g)]^{3/2} e^{3\alpha}$, where $\gamma$ is called the Immirzi parameter \cite{Immirzi} and $\Delta_g$ is the area gap (the minimum allowed nonzero eigenvalue of the area operator in LQG \cite{aps2}). The sign of $v$ depends on the triad orientation. In addition, the physical spatial volume is $V= 2\pi \gamma \Delta_g^{1/2} |v|$. The new variables $v$ and $b$ contain all the relevant information about the triad and the holonomies of the Asthekar-Barbero connection of the flat FLRW cosmology, within the improved dynamics scheme put forward in Ref. \cite{aps2}. While fluxes of the triad are functions of the volume $v$, the interesting holonomies have elements that are complex exponentials of $\pm b/2$. The corresponding Hilbert space $\mathcal{H}_{\text{kin}}^{\text{grav}}$ where the FLRW geometry is represented admits a basis of eigenstates of the volume, provided with the discrete inner product. For each of the states of this basis, the basic holonomy elements simply shift by one the eigenvalue of $v$. On the other hand, for the homogeneous scalar field we choose the much simpler space of square-integrable functions over the real line, $\mathcal{L}^2(\mathbb{R},d\phi)$, as the Hilbert space for a standard Schrödinger representation, with canonical variables $\{\phi,\pi_\phi\}$.

For the quantum representation of the gauge-invariant scalar and tensor perturbations, including their contribution to the zero mode of the Hamiltonian constraint, we adopt suitable Fock representations (for more details, see Refs. \cite{CMM,msq,tensor}). As for the quantum implementation of the linear perturbative constraints, essentially what they imply is that the physical states do not depend on perturbative gauge degrees of freedom. Finally, for the fermionic degrees of freedom, we choose a Fock representation $\mathcal{F}_D$ associated with creation and annihilationlike variables that have an asymptotic behavior in the sector of large $\omega_k$ determined by our previous considerations, and hence given by Eqs. \eqref{relacio}, \eqref{eq: ansatz1}, \eqref{eq: ansatz2}, and \eqref{eq: recursive}. These annihilationlike variables are promoted to the annihilation operators $\hat{a}^{(x,y)}_{\vec{k}}$ and $\hat{b}^{\ (x,y)}_{\vec{k}}$ for particle and antiparticle excitations, respectively, while the creationlike variables are represented by the adjoints of these operators. In the following, we denote this adjoint operation with a dagger. 

The complete system formed by all the sectors is subject to the zero mode of the Hamiltonian constraint, which we represent quantum mechanically with some extra prescriptions, additional to the ones fixed by the representation of the elementary variables. In particular, we adopt normal ordering for the products of creation and annihilation operators. The rest of the prescriptions refer to the representations of the functions of the FLRW geometry that appear as coefficients of the quadratic perturbative contributions to the constraint. Since we do not need them explicitly in the rest of our discussion, we refer the reader to Refs. \cite{fermihlqc,msq} for details about these prescriptions. 

We impose the zero mode of the Hamiltonian constraint \`a la Dirac \cite{Dirac}, with physical states annihilated by its (adjoint) action. Following the strategy of Refs. \cite{CMM,fermihlqc,msq}, we choose an ansatz with separation of variables: The wave functions of the physical states of interest factorize into partial wave functions that depend each on a different sector of the system, namely the FLRW geometry, the gauge-invariant scalar and tensor perturbations, and the fermionic perturbations. We allow that all these partial wave functions depend on the inflaton, which in this way will play the role of a relational time. Additionally, we ask that the part of the wave function that contains the geometric degrees of freedom, which we call $\Gamma(V,\phi)$, is normalized (in the discrete inner product for the volume) and has a unitary evolution in $\phi$ generated by a positive operator $\hat{\tilde{\mathcal{H}}}_0$, so that
\begin{equation}\label{eq: ing1}
-i\partial_\phi \Gamma(V,\phi)= \hat{\tilde{\mathcal{H}}}_0 \Gamma(V,\phi).
\end{equation} Finally, we restrict our attention to generators for which the action of $\partial_\phi^2+ \hat{\tilde{\mathcal{H}}}_0^2$  on $\Gamma$ differs from the corresponding action of the constraint of the unperturbed FLRW cosmology at most in a quadratic contribution of the perturbations. In this way, we contemplate the possibility that there exists some kind of quantum backreaction between the perturbations and the homogeneous background.

Furthermore, if we can ignore the transition between states of the FLRW geometry that are mediated by the action of our Hamiltonian constraint, all relevant information about this constraint can be captured by replacing its operator dependence on the homogeneous geometry with expectation values $\langle\cdot\rangle_\Gamma$ on the considered state $\Gamma$ in $\mathcal{H}_{\text{kin}}^{\text{grav}}$, computed with the discrete inner product. With this approximation and a kind of Born-Oppenheimer one, the imposition of the entire Hamiltonian constraint leads in fact to a set of Schr\"odinger equations, one for each of the different perturbative sectors of the system \cite{fermihlqc,CMM}. For the partial wave function that depends on the fermionic degrees of freedom, which we call $\psi(\mathcal{N}_D,\phi)$,\footnote{We use $\mathcal{N}_D$ as an abstract notation for the occupation numbers of the fermionic particles and antiparticles in the chosen Fock representation.} we arrive at the equation
\begin{equation}\label{eq: schrodinger}
-i \partial_\phi\psi_D(\mathcal{N}_D,\phi)= \frac{l_0 \langle \widehat{V^{2/3}e^{\alpha}\breve H_D}\rangle_\Gamma-C_D^{(\Gamma)}}{\langle \hat{\tilde{\mathcal{H}}}_0\rangle_\Gamma}\ \psi_D(\mathcal{N}_D,\phi)\equiv \mathcal{H}_D^{(\Gamma)}(\phi)\psi_D(\mathcal{N}_D,\phi).
\end{equation}
Here, the hat above a function of the FLRW geometry stands for its representation as a quantum operator. Besides, we have defined $\mathcal{H}_D^{(\Gamma)}(\phi)$, which can be regarded as a time-dependent (i.e., $\phi$-dependent) effective fermionic Hamiltonian operator that acts on $\mathcal{F}_D$ and generates the Schr\"odinger evolution. On the other hand, the term $C_D^{(\Gamma)}$, plus some similar terms in the Schr\"odinger equations of the gauge-invariant scalar and tensor perturbations, equals the expectation value in $\Gamma$ of the difference between the action of $\partial_\phi^2+\hat{\tilde{\mathcal{H}}}_0^2$ and the action of the Hamiltonian constraint of the unperturbed model. It is in this sense that we can call $C_D^{(\Gamma)}$ the fermionic backreaction, as it measures, in average, how much the homogeneous part $\Gamma$ of the solutions of the perturbed model departs from an unperturbed solution \cite{fermihlqc,backreaction}.

Let us now introduce a change to the time \begin{equation}d\eta_\Gamma=\frac{l_0 \langle \hat{V}^{2/3} \rangle_\Gamma}{\langle \hat{\tilde{\mathcal{H}}}_0\rangle_\Gamma}d\phi,\end{equation} which is well defined because $\tilde{\mathcal{H}}_0$ is positive and $\hat{V}$ has a strictly positive lower bound (at least in the adopted representation: see Ref. \cite{mmo}). It should be noted that, if we further restrict $\Gamma(V,\phi)$ to be highly peaked on classical or effective trajectories, this time coincides (up to corrections that are quadratic in perturbations) with the standard conformal time in cosmology as far as we circumscribe it within an interval where the inflaton is monotonic. Nonetheless, our definition of $\eta_{\Gamma}$  is perfectly consistent in the quantum theory beyond such a classical or effective regime, provided that the involved expectation values remain strictly positive and finite. In this sense, $\eta_{\Gamma}$ may be regarded as a relational time for the different parts of the wave function, and one should only attempt to find a correspondence with a standard conformal time within regimes and intervals of the type that we have commented.

Using then Eq. \eqref{eq: schrodinger} and the definition of our creation and annihilationlike variables, we obtain the following Heisenberg relations (in the considered asymptotic regime of large $\omega_k$), evaluated at $\eta_\Gamma=\eta$:
\begin{align}\label{eq: Heisenberg}
d_{\eta_\Gamma}\hat{a}^{(x,y)}_{\vec{k}}(\eta,\eta_0)&=-iF^{(\Gamma)}_{\vec{k}}\hat{a}^{(x,y)}_{\vec{k}}(\eta,\eta_0), \quad
d_{\eta_\Gamma}\hat{b}^{\dagger(x,y)}_{\vec{k}}(\eta,\eta_0)=i\left(F^{(\Gamma)}_{\vec{k}}+J^{(\Gamma)}_{\vec{k}}\right)\hat{b}^{\dagger(x,y)}_{\vec{k}}(\eta,\eta_0),
\end{align}
where we have called
\begin{align}
F^{(\Gamma)}_{\vec{k}}=\frac{2\langle \widehat{V^{2/3}e^{\alpha} h^{\vec{k},(x,y)}_D}\rangle_\Gamma}{\langle \widehat{V}^{2/3}\rangle_\Gamma}, \quad J^{(\Gamma)}_{\vec{k}}=\frac{2\langle \widehat{V^{2/3}e^{\alpha} h^{\vec{k},(x,y)}_J}\rangle_\Gamma}{\langle \widehat{V}^{2/3}\rangle_\Gamma}.
\end{align}
We can integrate these Heisenberg equations to obtain, asymptotically, the following Bogoliubov transformation
\begin{align}\label{bog}
\hat{a}^{(x,y)}_{\vec{k}}(\eta,\eta_0)= e^{-iF_{\eta,\vec{k}}^{(\Gamma)}}\hat{a}^{(x,y)}, \quad
\hat{b}^{\dagger(x,y)}_{\vec{k}}(\eta,\eta_0)= e^{i\left(F_{\eta,\vec{k}}^{(\Gamma)}+J_{\eta,\vec{k}}^{(\Gamma)}\right)}\hat{b}^{\dagger(x,y)},
\end{align}
where we have defined $\hat{a}^{(x,y)}_{\vec{k}}(\eta_0)=\hat{a}^{(x,y)}_{\vec{k}}$, $\hat{b}^{\dagger(x,y)}_{\vec{k}}(\eta_0)=\hat{b}^{\dagger(x,y)}_{\vec{k}}$ and 
\begin{align}
F_{\eta,\vec{k}}^{(\Gamma)}=\int_{\eta_0}^\eta F_{\vec{k}}^{(\Gamma)} d\eta_\Gamma,\quad J_{\eta,\vec{k}}^{(\Gamma)}=\int_{\eta_0}^\eta J_{\vec{k}}^{(\Gamma)}d\eta_\Gamma.
\end{align}
This Bogoliubov transformation is clearly unitary \cite{shale}, because it does not mix creation and annihilation operators in the considered asymptotic region, so that the antilinear part of the transformation vanishes asymptotically (or, more precisely, can be made of an asymptotic order as negligible as desired). 

Finally, we can construct an operator $\hat{T}$ defined as
\begin{align}
\hat T=\sum_{\vec{k},(x,y)}\hat{T}_{\vec{k}}^{(x,y)},\qquad \hat{T}_{\vec{k}}^{(x,y)}=i F_{\eta,\vec{k}}^{(\Gamma)}\left(\hat{a}^{\dagger(x,y)}_{\vec{k}}\hat{a}^{(x,y)}_{\vec{k}}+\hat{b}^{\dagger(x,y)}_{\vec{k}}\hat{b}^{(x,y)}_{\vec{k}}
\right)+i J_{\eta,\vec{k}}^{(\Gamma)}\left(\hat{b}^{\dagger(x,y)}_{\vec{k}}\hat{b}^{(x,y)}_{\vec{k}}\right),
\end{align}
such that in the regime of large $\omega_k$
\begin{align}
e^{\hat T}\hat{a}^{(x,y)}_{\vec{k}}e^{-\hat T}&= \hat{a}^{(x,y)}_{\vec{k}}(\eta,\eta_0),\\ 
e^{\hat T} \hat{b}^{\dagger(x,y)}_{\vec{k}} e^{-\hat T}&= \hat{b}^{\dagger(x,y)}_{\vec{k}}(\eta,\eta_0),
\end{align}
as can be checked using Hadamard's lemma \cite{hadammardlemma}. Hence, $e^{-\hat T}$ can be taken as the unitary operator that implements the dynamical Bogoliubov transformation \eqref{bog}.

This confirms that, asymptotically, the vacuum (namely the normalized state in the kernel of all the annihilation operators) is stationary under the evolution dictated by this operator. In turn, this implies that the vacuum is annihilated by the left-hand side of Eq. \eqref{eq: schrodinger}, up to possibly the contribution of a complex phase. And, since we adopted normal ordering for the representation of the zero mode of the Hamiltonian constraint, the first term of the right-hand side of Eq. \eqref{eq: schrodinger} annihilates the vacuum as well. Hence, with our choice of Fock representation, the fermionic backreaction $C_D^{(\Gamma)}$ for the vacuum (which was only convergent in Ref. \cite{fermihlqc} after a subtraction of infinities scheme) is not only finite now, but indeed can straightforwardly be set to vanish, at least asymptotically. In this way, the vacuum remains invariant in the evolution. Then, it is clear that its image under the action of the fermionic Hamiltonian is a normalizable state in $\mathcal{F}_D$. As a consequence, we conclude that the fermionic Hamiltonian is properly defined in the dense subset of $\mathcal{F}_D$ spanned by the $n$-particle/antiparticle states with a finite number of fermionic excitations.

\section{Conclusions}\label{conclusiones}

We have investigated the choice of a vacuum for the Dirac field in an inflationary flat FLRW spacetime by suplementing with extra physical requirements the criterion of invariance under spatial symmetries and unitary Heisenberg evolution that has been explored in the literature recently. More specifically, the additional requirement that we have considered is the diagonalization of the fermionic contribution to the Hamiltonian constraint of the gravitational system when the fermions are treated as perturbative fields on an average (possibly quantum mechanically dressed) background, in the asymptotic regime of large wave numbers, that we identify with the eigenvalues of the Dirac operator on the spatial sections of the cosmology. While the original criterion of spatial symmetry invariance and dynamical unitarity leads to a family of Fock representations that are unitarily equivalent among them, but still leaving an infinite freedom in the choice of a vacuum, the inclusion of the diagonalization requirement has been  shown to restrict the available freedom to the choice of two complex phases and terms that are negligible at any desired order in an expansion in inverse powers of the wave number.

In our analysis, we have truncated the Einstein-Dirac action, also in the presence of a scalar field that plays the role of an inflaton, at quadratic order in perturbations, treating as such the nonzero Fourier modes of the metric and the scalar field and all the contributions of the Dirac field. The zero modes of the inflaton and the metric have been treated exactly. Because of this, the action does not get contributions that are linear in our perturbations, including those of the lapse and the shift (actually, this statement holds beyond our quadratic truncation, in higher-order perturbative schemes). Apart from linear perturbative constraints, the system is subject to the zero mode of the Hamiltonian constraint, which contains a term that is formally identical to the global Hamiltonian constraint of the unperturbed model, but in addition includes other contributions that are quadratic in the perturbations, in particular a fermionic term. These perturbations can be described by (an Abelianized version of) the linear perturbative constraints, gauge variables that are momenta of those constraints, and gauge invariants that commute with all the former quantities. For the metric and the inflaton, one can choose as gauge invariants the Mukhanov-Sasaki field and the tensor perturbations, together with their canonical momenta. The fermionic perturbations, on the other hand, are immediately gauge invariants, because the Einstein-Dirac action is quadratic in the Dirac field, so that the latter couples directly to the unperturbed tetrad when we truncate the action in our scheme. The above set of perturbative variables can be completed into a canonical set for the whole system, at the considered truncation order, by including zero modes that are suitably corrected with quadratic perturbative terms \cite{CMM}.

Focusing our discussion on the fermionic sector of the inhomogeneities, one still has considerable liberty in the way in which one can separate the fermionic degrees of freedom from those of the homogeneous background through canonical transformations, with a splitting that maintains the gauge-invariant character of the fermionic variables. Different splittings of the fermionic and the zero-mode sectors of phase space result in different quantum behaviors for the combined system and different quantum dynamics for the fermionic variables, since the separation amounts to a background-dependent (and hence dynamical) redefinition of the basic, creation and annihilationlike fermionic variables. Instead of regarding this ambiguity as a complication, we have taken advantage of it and looked for a choice of those creation and annihilationlike variables such that the part of the Hamiltonian constraint that rules their evolution has good physical properties. In order to do this, we have considered all possible linear combinations of the fermionic mode coefficients that define creation and annihilationlike variables, allowing these combinations to depend on zero modes. Among all the viable combinations that do not mix fermionic modes, and therefore respect the spatial symmetries of the model, we have then sought for choices that lead to a especially simple fermionic Hamiltonian, without interaction terms that would create and destroy pairs of particles and antiparticles (at least in the ultraviolet sector of large wave numbers). The absence of these interactions amounts to the (asymptotic) diagonalization of the fermionic contribution to the zero mode of the Hamiltonian constraint. We have shown that it is possible to attain this diagonalization at any asymptotic order in inverse powers of the wave number, and that the resulting characterization of fermionic variables is unique up to certain phases at that order and up to terms that are negligible in the asymptotic series of the coefficients that define the creation and annihilationlike variables.

Combining then this Fock representation for the Dirac field, suitable Fock representations for the rest of gauge invariants, and an LQG-inspired quantization for the homogeneous sector, we have considered the hybrid quantization of the system. Given our choice of fermionic creation and annihilation variables, the Fock representation determined by them has a considerably simple quantum dynamics. We have shown this in detail by adopting an ansatz with separation of variables for the physical wave functions and introducing a kind of Born-Oppenheimer approximation in which we neglected changes in the FLRW geometry mediated by the zero mode of the Hamiltonian constraint. The relevant information about this constraint is then captured in its expectation value on the partial wave function that describes the background FLRW cosmology. In this manner, one obtains a Schr\"odinger equation for the fermionic degrees of freedom that, with our (asymptotic) choice of vacuum, leads to Heisenberg equations that do not mix the creation and annihilation operators of the particles and antiparticles. Thus, the dynamical Bogoliubov transformation of these operators can be implemented trivially as a quantum unitary transformation in the discussed asymptotic regime, because its antilinear part can be made equal to $0$ at any desired asymptotic order. We also have constructed an evolution operator that implements this Bogoliubov transformation, and checked that it leaves the vacuum stationary. This property and the fact that the fermionic contribution to the zero mode of the Hamiltonian constraint annihilates the vacuum guarantee that the backreaction term in the Schr\"odinger equation not only does not diverge, but can be made negligible at any asymptotic order in inverse powers of the wave number, without the need of any regularization scheme.

Let us emphasize that our choice of annihilation and creationlike variables, which leads to an asymptotically diagonal fermionic Hamiltonian, is carried out within a canonical framework that is valid for the entire, truncated, cosmology, with a phase space that describes not only the fermionic perturbations, but also the background variables.
In particular, the choice provides a specific splitting between the two sectors of the cosmological system: the homogeneous one and the perturbations. Besides, since the zero mode of the Hamiltonian constraint for this entire cosmology contains a term quadratic in all of the perturbations, the classical dynamics of the selected fermionic variables is only linear if one ignores their backreaction on the homogeneous sector. Nevertheless, the resulting Hamiltonian constraint applies to the entire cosmology even beyond this linearized context, keeping its nice properties for quantization even in that extended scenario. Furthermore, we have seen that it is possible to reach a regime in the quantum dynamics of the entire cosmology where the fermionic wave function effectively obeys a Schr\"odinger equation. The \emph{effective} Hamiltonian operator that drives this evolution actually corresponds to the Fock representation of the fermionic contribution to the constraint, but with its background dependence replaced with expectation values in the partial wave function that describes the homogeneous geometry. In this sense, the hybrid quantum theory allows for a generalized linearized regime of a fermionic field that propagates over a mean quantum background, which need not follow any classical or effective trajectory and might even allow for some backreaction effects.

Our conclusions support some aspects of a similar study under development for the case of cosmological scalar perturbations in flat FLRW cosmologies, where the requirement of Hamiltonian diagonalization appears to supplement satisfactorily the criterion of spatial symmetry invariance and unitarity of the Heisenberg evolution \cite{bgt}. It would be interesting to discuss other aspects explored in that work about the properties of the selected vacuum and its relation with adiabatic states, taking into account the results that are known about such states for fermionic fields \cite{hollands} and extending them with further investigations. 

\acknowledgments

The authors are grateful to J. Cortez, T. Thiemann, and J.M. Velhinho for enlighting conversations. This work was supported by Grants No. MINECO FIS2014-54800-C2-2-P and No. MINECO FIS2017-86497-C2-2-P from Spain.


\begin{thebibliography}{0}

\bibitem{Stone} M.H. Stone, {\it On one-parameter unitary groups in Hilbert Space}, Ann. Math. {\bf 33}, 643 (1932).
  
\bibitem{VN} G.B. Folland, {\it Harmonic Analysis in Phase Space} (Princeton University Press, Princeton, 1989).
  
\bibitem{Peskin} M.E. Peskin and D.V. Schroeder, {\it Introduction to quantum field theory} (Perseus Books, Reading, 1995).

\bibitem{haag} R. Haag, \textit{Local Quantum Physics} (Springer, New York, 1996).
  
\bibitem{energy1} A. Ashtekar and A. Magnon, {\it Quantum fields in curved spacetimes}, Proc. R. Soc. A {\bf 346}, 375 (1975).
  
\bibitem{energy2} A. Ashtekar and A. Magnon-Ashtekar, {\it A curiosity concerning the role of coherent states in quantum field theory}, Pramana {\bf 15}, 1071980 (1980). 
  
\bibitem{Wald} R.M. Wald, {\it Quantum field theory in Curved Spacetimes and Black Hole Thermodynamics} (University of Chicago Press, Chicago, 1994).

\bibitem{bidav} N.D. Birrell and P.C.W. Davies,  {\it Quantum Fields in Curved Space} (Cambridge University Press, Cambridge, England, 1982).
  
\bibitem{uniqueness11} J. Cortez, G.A. Mena Marug\'an, and J.M. Velhinho, {\it Uniqueness of the Fock quantization of the Gowdy T3 model}, Phys. Rev. D {\bf 75}, 084027 (2007).
   
\bibitem{uniqueness10} J. Cortez, G.A. Mena Marug\'an,  and J.M. Velhinho, {\it Fock quantization of a scalar field with time-dependent mass on the three-sphere: Unitarity and uniqueness}, Phys. Rev. D {\bf 81}, 044037 (2010).
    
\bibitem{uniqueness9} M. Fern\'andez-M\'endez, G.A. Mena Marug\'an, J. Olmedo, and J.M. Velhinho, {\it Unique Fock quantization of scalar cosmological perturbations}, Phys. Rev. D {\bf85}, 103525 (2012).
  
\bibitem{uniqueness8} L. Castell\'o Gomar, J. Cortez, D. Mart\'{\i}n-de Blas, G.A. Mena Marug\'an, and J.M. Velhinho, {\it Uniqueness of the Fock quantization of scalar fields in spatially flat cosmological spacetimes}, J. Cosmol. Astropart. Phys. {\bf 11} (2012) 001.

\bibitem{uniqueness7} L. Castell\'o Gomar, J. Cortez, D. Mart\'{\i}n-de Blas, G.A. Mena Marug\'an, and J.M. Velhinho, {\it Unitary evolution and uniqueness of the Fock quantization in flat cosmologies with compact spatial sections}, Electron. J. Theor. Phys. {\bf 11}, 43 (2014).

\bibitem{uniqueness6} J. Cortez, D. Mart\'{\i}n-de Blas, G.A. Mena Marug\'an, and J.M. Velhinho, {\it Massless scalar field in de Sitter spacetime: Unitary quantum time evolution}, Classical Quantum Gravity {\bf 30}, 075015 (2013).
  
\bibitem{uniqueness5} G.A. Mena Marugán, D. Mart\'{\i}n-de Blas, and L. Castell\'o Gomar, {\it Unitary evolution and uniqueness of the Fock quantization in flat cosmologies}, J. Phys. Conf. Ser. {\bf 410}, 012151 (2013).
   
\bibitem{uniqueness4} L. Castelló Gomar and  G. A. Mena Marugán, \textit{Uniqueness of the Fock quantization of scalar fields and processes with signature change in cosmology}, Phys. Rev. D {\bf 89}, 084052 (2014).
  
\bibitem{uniqueness3} J. Cortez, G.A. Mena Marug\'an,  and J.M. Velhinho, {\it Quantum unitary dynamics in cosmological spacetimes}, Ann. Phys. (Amsterdam) {\bf 363}, 36 (2015).
  
\bibitem{uniqueness2} J. Cortez, B. Elizaga Navascu\'es, M. Mart\'{\i}n-Benito, G.A. Mena Marug\'an, and J.M. Velhinho, {\it Unitary evolution and uniqueness of the Fock representation of Dirac fields in cosmological spacetimes}, Phys. Rev. D {\bf 92}, 105013 (2015).
  
\bibitem{uniqueness1} J. Cortez, B. Elizaga Navascu\'es, M. Mart\'{\i}n-Benito, G.A. Mena Marug\'an, J. Olmedo, and J.M. Velhinho, {\it Uniqueness of the Fock quantization of scalar fields in a Bianchi I cosmology with unitary dynamics}, Phys. Rev. D {\bf 94}, 105019 (2016).
  
\bibitem{uniqfermi}J. Cortez, B. Elizaga Navascu\'es, M. Mart\'{\i}n-Benito, G.A. Mena Marug\'an, and J.M. Velhinho, {\it Dirac fields in flat FLRW cosmology: Uniqueness of the Fock quantization}, Ann. Phys. (Amsterdam) {\bf 376}, 76 (2017).

\bibitem{Thiemann} T. Thiemann, {\it Modern Canonical Quantum general relativity} (Cambridge University Press, Cambridge, England, 2007). 

\bibitem{hlqc1} M. Mart\'{\i}n-Benito, L.J. Garay, and G.A. Mena Marug\'an, {\it Hybrid quantum Gowdy cosmology: Combining loop and Fock quantizations}, Phys. Rev. D {\bf 78}, 083516 (2008).
  
\bibitem{hlqc2} G.A. Mena Marug\'an and M. Mart\'{\i}n-Benito, {\it Hybrid quantum cosmology: Combining Loop and Fock quantizations}, Int. J. Mod. Phys. A {\bf 24}, 2820 (2009).
  
\bibitem{hlqc3} M. Fern\'andez-M\'endez, G.A. Mena Marug\'an, and J. Olmedo, {\it Hybrid quantization of an inflationary universe}, Phys. Rev. D {\bf 86}, 024003 (2012).

\bibitem{backreaction} B. Elizaga Navascu\'es, G.A. Mena Marug\'an, and S. Prado Loy, {\it Backreaction of fermionic perturbations in the Hamiltonian of hybrid loop quantum cosmology}, Phys. Rev. D {\bf 98}, 063535 (2018).
  
\bibitem{fermihlqc} B. Elizaga Navascu\'es, M. Mart\'{\i}n-Benito, and G.A. Mena Marug\'an, {\it Fermions in hybrid loop quantum cosmology}, Phys.
Rev. D {\bf 96}, 044023 (2017).

\bibitem{Death} P.D. D’Eath and J.J. Halliwell, {\it Fermions in quantum cosmology}, Phys. Rev. D {\bf 35}, 1100 (1987).

\bibitem{Hawking} J.J. Halliwell and S.W. Hawking, {\it Origin of structure in the universe}, Phys. Rev. D {\bf 31}, 1777 (1985).

\bibitem{CMM} L. Castell\'o Gomar, M. Mart\'{\i}n-Benito, and G.A. Mena Marug\'an, {\it Gauge-invariant perturbations in hybrid quantum cosmology}, J. Cosmol. Astropart. Phys. {\bf 06} (2015) 045.

\bibitem{tensor} F. Ben\'{\i}tez Mart\'{\i}nez and J. Olmedo, {\it Primordial tensor modes of the early Universe}, Phys. Rev. D {\bf 93}, 124008 (2016).

\bibitem{Bardeen} J.M. Bardeen, {\it Gauge-invariant cosmological perturbations}, Phys. Rev. D {\bf 22}, 1882 (1980).

\bibitem{Mukhanov} V. Mukhanov, {\it Physical Foundations of Cosmology} (Cambridge University Press, Cambridge, England, 2005).

\bibitem{MS} V. Mukhanov, {\it Quantum theory of gauge-invariant cosmological perturbations}, Zh. Eksp. Teor. Fiz. {\bf 94}, 1 (1988) [Sov. Phys. JETP {\bf 67}, 1297 (1988)].

\bibitem{Sasaki} M. Sasaki, {\it Gauge invariant scalar perturbations in the new inflationary universe}, Prog. Theor. Phys. {\bf 70}, 394 (1983).

\bibitem{kodasasa} H. Kodama and M. Sasaki, {\it Cosmological perturbation theory}, Prog. Theor. Phys. Suppl. {\bf 78}, 1 (1984).

\bibitem{TN} J.E. Nelson and C. Teitelboim, {\it Hamiltonian formulation of the theory of interacting gravitational and electron fields}, Ann. Phys. (N.Y.) {\bf 116}, 86 (1978).

\bibitem{mmo} M. Mart\'{\i}n-Benito, G.A. Mena Marug\'an, and J. Olmedo, {\it Further improvements in the understanding of isotropic loop quantum cosmology}, Phys. Rev. D {\bf 80}, 104015 (2009).

\bibitem{aps1} A. Ashtekar, T. Pawlowski, and P. Singh, {\it Quantum nature of the big bang: An analytical and numerical investigation}, Phys. Rev. D {\bf 73}, 124038 (2006).

\bibitem{aps2} A. Ashtekar, T. Pawlowski, and P. Singh, {\it Quantum nature of the big bang: improved dynamics}, Phys. Rev. D {\bf 74}, 084003 (2006).

\bibitem{AshSingh} A. Ashtekar and P. Singh, {\it Loop quantum cosmology: A status report}, Classical Quantum Gravity {\bf 28}, 213001 (2011).

\bibitem{Immirzi} G. Immirzi, {\it Quantum gravity and Regge calculus}, Nucl. Phys. Proc. Suppl. {\bf 57}, 65 (1997).

\bibitem{msq} L. Castell\'o Gomar, M. Fern\'andez-M\'endez, G.A. Mena Marug\'an, and J. Olmedo, {\it Cosmological perturbations in Hybrid
loop quantum cosmology: Mukhanov–Sasaki variables}, Phys. Rev. D {\bf 90}, 064015 (2014).

\bibitem{Dirac} P.A.M. Dirac, {\it Lectures on quantum mechanics} (Belfer Graduate School of Science, New York, 1964).

\bibitem{shale} D. Shale, {\it Linear symmetries of free boson fields}, Trans. Am. Math. Soc. {\bf 103}, 149 (1962).

\bibitem{hadammardlemma} J. Nestruev, {\it Smooth Manifolds and Observables} (Springer, New York, 2002).

\bibitem{bgt} B. Elizaga Navascu\'es, G.A. Mena Marug\'an, and T. Thiemann, {\it Hamiltonian diagonalization in hybrid quantum cosmology}, arXiv:1903.05695.

\bibitem{hollands} S Hollands, {\it The Hadamard condition for Dirac fields and adiabatic states on Robertson-Walker spacetimes},  	Commun. Math. Phys. {\bf 216}, 635 (2001).
  
  
\end{thebibliography}
\end{document}